\documentclass[sigconf]{acmart}

\usepackage[english]{babel}
\usepackage{CJKutf8}
\usepackage{xurl}
\usepackage{xspace}
\usepackage{listings}

\usepackage{amsmath,amsfonts,bm}

\def\eqref#1{equation~\ref{#1}}

\def\1{\bm{1}}

\DeclareMathAlphabet{\mathsfit}{\encodingdefault}{\sfdefault}{m}{sl}
\SetMathAlphabet{\mathsfit}{bold}{\encodingdefault}{\sfdefault}{bx}{n}

\AtBeginEnvironment{itemize}{\setlength{\leftmargini}{1.5em}}

\newcommand{\parabf}[1]{\smallskip\noindent\textbf{#1}}
\newcommand{\sysname}{KnitTrans\xspace}

\definecolor{lightblue}{RGB}{230,240,255}
\definecolor{redlinkcolor}{rgb}{0.79607843,0.25098039,0.25882353}
\definecolor{bluecitecolor}{rgb}{0,0.36,0.69}
\hypersetup{colorlinks=true,linkcolor=redlinkcolor,citecolor=bluecitecolor,urlcolor=bluecitecolor}

\renewcommand\footnotetextcopyrightpermission[1]{}
\setcopyright{none}
\acmDOI{}
\acmISBN{}
\acmPrice{}
\acmYear{}
\copyrightyear{}
\acmConference{}{}{}
\acmBooktitle{}
\title[\sysname: Coordinated Trajectory Transfer for LLM Reinforcement Learning]{\sysname: Coordinated Trajectory Transfer for Effective and
  Efficient LLM Reinforcement Learning}

\author{\texorpdfstring{%
  \normalfont\fontsize{11}{15}\selectfont
  \begin{tabular}{@{}cccc@{}}
    Bangjun Xiao$^{\dagger*}$ &
    Chiheng Lou$^{\dagger*}$ &
    Xiaodong Ji$^{\ddagger}$ &
    Xiangwei Deng$^{\ddagger}$ \\
    Yikai Zhao$^{\ddagger}$ &
    Gang Huang$^{\dagger}$ &
    Yun Ma$^{\dagger}$ &
    Hailin Zhang$^{\ddagger\diamond}$
  \end{tabular}%
}{Bangjun Xiao, Chiheng Lou, Xiaodong Ji, Xiangwei Deng,
  Yikai Zhao, Gang Huang, Yun Ma, Hailin Zhang}}
\affiliation[obeypunctuation=true]{%
  \normalfont\fontsize{11}{15}\selectfont\itshape
  \institution{$^{\dagger}$School of Computer Science, Peking University}
  \qquad
  \institution{$^{\ddagger}$LLM-Core Xiaomi}
}
\renewcommand{\shortauthors}{Bangjun Xiao et al.}
\thanks{$^{*}$Work done during internship at Xiaomi Corporation.}
\thanks{$^{\diamond}$Corresponding author.}

\begin{document}

\begin{abstract}
Trajectory transfer from rollout to training affects both
learning quality and system efficiency in large language model
reinforcement learning (RL). Enforcing batch-wide training
requirements can delay the admission of completed trajectories,
resulting in higher architectural staleness---the additional
parameter-version gap from disaggregated pipelining
relative to colocated execution. Beyond admission,
misalignment between data access and training consumption
causes redundant movement and unnecessary waiting when transferring
large trajectory payloads.

We present \sysname, a system that coordinates the
rollout-to-training path through three designs.
First, constraint-preserving mini-batch admission allows
flexibility in early mini-batch composition, while
checking that enough capacity remains to satisfy the remaining
quotas. Second, on-demand trajectory data access matches storage
units to consumer assignments and overlaps micro-batch data
preparation with computation. Third, multimodal encoding
deduplication supports reuse across a mini-batch while preserving
gradient contributions from all backbone micro-batches.
Experiments show up to 93\% lower architectural
staleness than with train-batch admission, narrowing the learning-quality
gap between disaggregated execution and colocated execution. On multimodal workloads, \sysname
reduces transfer latency by two orders of magnitude and
achieves a speedup of up to $5.8\times$ in transfer
plus training over the baseline.
\end{abstract}

\maketitle

\section{Introduction}
\label{sec:intro}

Reinforcement learning (RL) has become a pivotal component in the
post-training of large language models (LLMs), used to align model
behavior with human preferences~\cite{ouyang2022training} and to
elicit advanced capabilities such as
reasoning~\cite{openai2024o1,guo2025deepseek}, tool
use~\cite{jin2025searchr1,feng2025retool}, and multimodal
reasoning~\cite{huang2025visionr1,kimi2026k25}. Despite the numerous
RL algorithm
variants~\cite{schulman2017proximal,shao2024deepseekmath,yu2025dapo},
the workflow remains invariant: rollout generates trajectories;
after rewards are computed, training consumes them to update the
policy, which in turn generates subsequent
trajectories.
As RL applications advance~\cite{kimi2026k3,datamix,glm2026glm5,deepseek2026v4}
and training scales to larger batch sizes, larger clusters, and more
complex workloads, effective and efficient execution depends not only on rollout
and training individually, but also on trajectory transfer from rollout
to training. This handoff determines which trajectories enter training,
when they become available, how their data are stored, and
how those data are transferred to training workers.
Along this path, we focus on two aspects to improve learning
quality and execution efficiency, respectively: enabling early
admission under batch-wide training requirements, and aligning
trajectory data access with the spatial and temporal needs of
distributed training.

\begin{figure}[t]
  \centering
    \includegraphics[width=\linewidth]{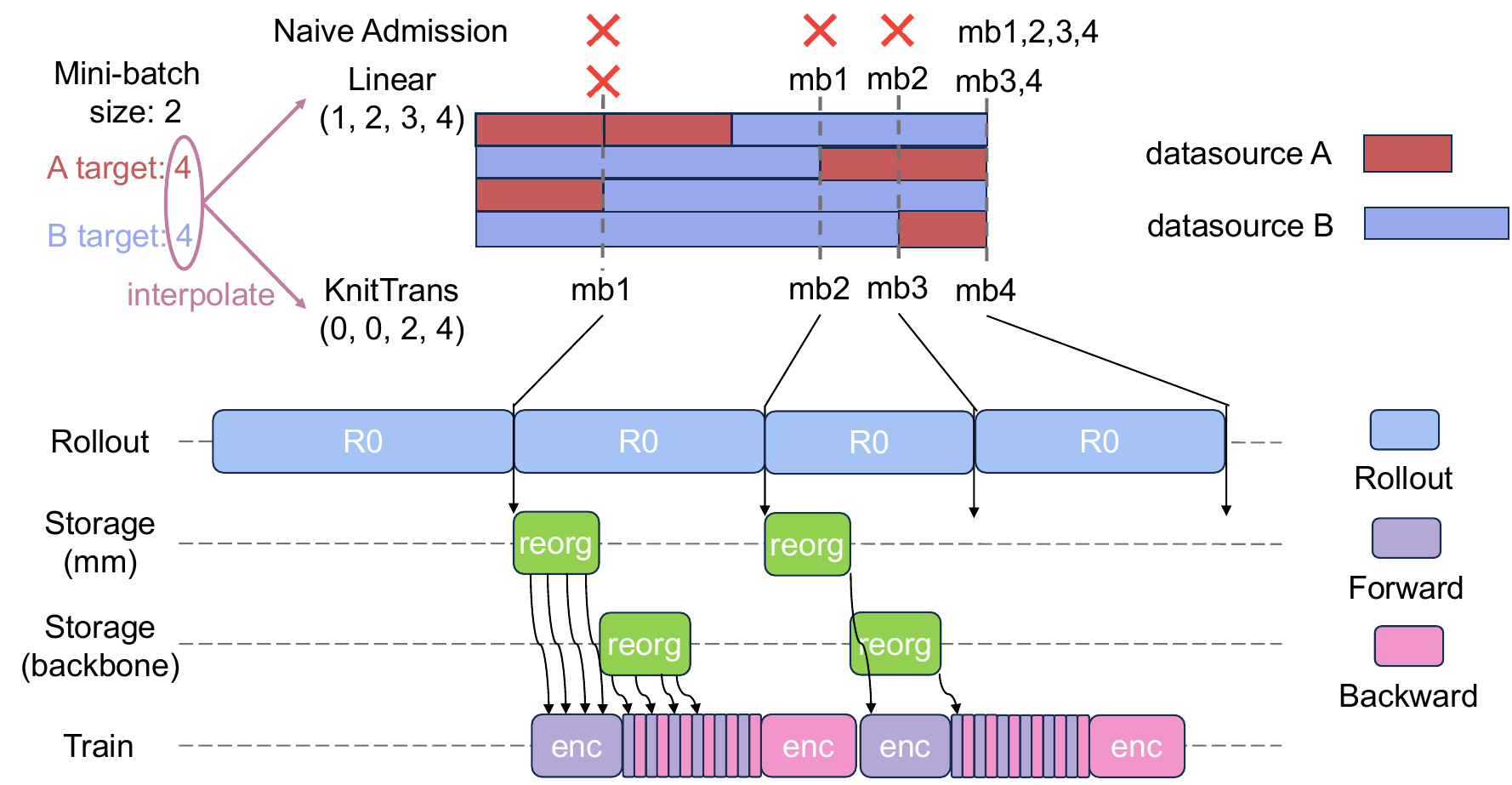}
  \par\vspace{-0.1in}
  \caption{\label{fig:intro} Timeline of trajectory admission,
    data delivery, and training computation under \sysname.}
  \vspace{-0.2in}
\end{figure}

RL training can impose batch-wide requirements: mixed RL
enforces target data-source ratios~\cite{glm2026glm5,datamix},
while partial-rollout completion quotas require sufficient
progress from earlier rollout steps to mitigate long-tail
starvation. Fine-grained streaming
~\cite{zhong2025streamrl,han2025asyncflow} makes completed
trajectories incrementally available, but admitting too many
from faster sources can leave insufficient capacity to satisfy
slower sources' quotas. Enforcing these requirements through
whole-train-batch admission creates a \emph{train-batch admission
barrier}: training waits until all admission conditions hold,
even when some trajectories are already ready
(Figure~\ref{fig:intro}). In off-policy disaggregated execution,
this delays training and parameter feedback to rollout,
resulting in higher \emph{architectural staleness}---the
additional parameter-version gap introduced by pipelining
relative to colocated execution, even under identical
staleness-control settings. This additional staleness degrades
learning quality.
The challenge is therefore to overcome the train-batch admission
barrier while preserving batch-wide requirements
(\S\ref{sec:motivation-staleness}).

Large trajectory payloads can also prolong training steps with
inefficient trajectory transfer. Long trajectories carry not only
token IDs but also per-token auxiliary data such as routing
decisions~\cite{r3} and top-$p$ sampling masks~\cite{mai_thinking_1},
alongside multimodal inputs such as images and
videos~\cite{mmrl,kimi2026k25}. Unlike pre-training on static datasets,
RL consumes trajectories generated online by rollout, 
in both colocated and disaggregated execution, placing
their preparation and transfer on the dependency path to policy
updates.
Spatially, storage organization determines
which data can be accessed independently, while the training
topology determines which data each worker needs.
Tensor-parallel (TP) ranks share inputs, context-parallel (CP)
ranks consume sequence slices, and multimodal encoders can have
assignments distinct from the backbone's. Existing sample-level
storage layouts~\cite{han2025asyncflow,mooncake2026miles} organize
data around sample boundaries rather than these consumer-specific
requirements, causing redundant reads or transfer of unneeded
data. Temporally, training consumes an admitted mini-batch
through a sequence of micro-batches. Existing RL systems
~\cite{sheng2025hybridflow,thudm2025slime} serialize payload
retrieval and input packing with computation on the critical
path, so these preparation costs prolong training time even
without redundant data movement.
The challenge is therefore to match both consumer-specific
data requirements and micro-batch consumption order, avoiding
redundant movement and unnecessary waiting
(\S\ref{sec:motivation-storage}).

We present \emph{\sysname}, which coordinates trajectory admission,
data access, and multimodal encoding along the rollout-to-training
path through three designs.

\parabf{Constraint-Preserving Mini-Batch Admission.}
\sysname designs progressive readiness targets that accommodate
uneven rollout progress, rather than require early admissions to
mirror the final train-batch composition. In Figure~\ref{fig:intro},
linear source-readiness targets delay the first mini-batch until
the slower source makes sufficient progress, whereas relaxed
early targets permit earlier admission.
To keep this flexibility from making the final train batch
infeasible, \emph{feasibility guards} check that each admission
leaves enough batch capacity to satisfy the remaining
batch-composition quotas. 

\parabf{On-Demand Trajectory Data Access.}
\sysname aligns storage and data access with training consumption
in space and time. It stores trajectory data as independently
addressable atomic units and reorganizes the required units
according to the training topology, reducing redundant data
movement. After admission, data preparation and delivery
proceed incrementally in micro-batch consumption order and
overlap with training computation, as illustrated in
Figure~\ref{fig:intro}. 

\parabf{Multimodal Encoding Deduplication.}
GRPO-style RL~\cite{shao2024deepseekmath} produces multiple
trajectories sharing the same multimodal inputs. Sharing storage
does not itself remove repeated encoding of these inputs during
training. \sysname encodes each unique multimodal input once per
mini-batch, reuses its encoding across the consuming trajectories,
and accumulates their gradients for the encoder update. 

\smallskip\noindent We evaluate \sysname on a 256-GPU cluster using models with 235B–671B 
total parameters, covering agentic and multimodal RL workloads under both 
disaggregated and colocated architectures. 
In disaggregated training, \sysname reduces architectural staleness
by up to 93\% relative to train-batch admission, narrowing the 
learning-quality gap with colocated training.
In a separate evaluation of constraint preservation,
\sysname satisfies all data-source quotas and
reduces architectural staleness by 39.2\% compared with linear-target
admission. On multimodal workloads, it reduces exposed transfer
latency by two orders of magnitude and speeds up transfer plus
training by $5.8\times$ over TransferQueue.

We list the contributions of this paper as follows.
\begin{itemize}
\item We analyze how train-batch admission barriers under
batch-wide requirements delay consumption and increase
architectural staleness, and how mismatches between data access
and training consumption cause redundant movement and unnecessary
waiting.

\item We develop constraint-preserving mini-batch admission using
progressive readiness targets and feasibility guards, and on-demand
trajectory data access through atomic-unit storage, topology-aware
reorganization, and consumption-driven delivery. Mini-batch-wide
multimodal encoding reuse further reduces repeated computation and
accelerates training.

\item We build \sysname, an RL system that realizes these designs
across disaggregated and colocated architectures. Large-scale evaluation
demonstrates substantial gains in both learning quality and training
efficiency through improvements to the rollout-to-training path.
\end{itemize}

\begin{figure*}[t]
  \centering
    \begin{minipage}[c]{0.38\linewidth}
      \centering
      \includegraphics[width=\linewidth]{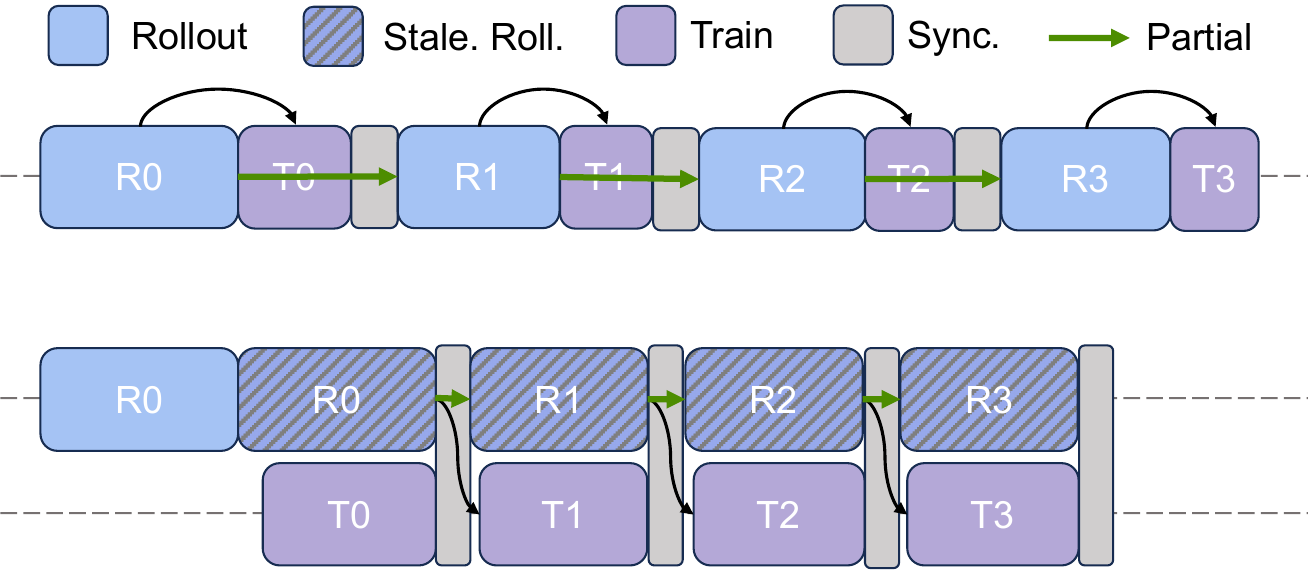}
    \end{minipage}%
    \hspace{0.02\linewidth}%
    \begin{minipage}[c]{0.45\linewidth}
      \centering
      \includegraphics[width=\linewidth]{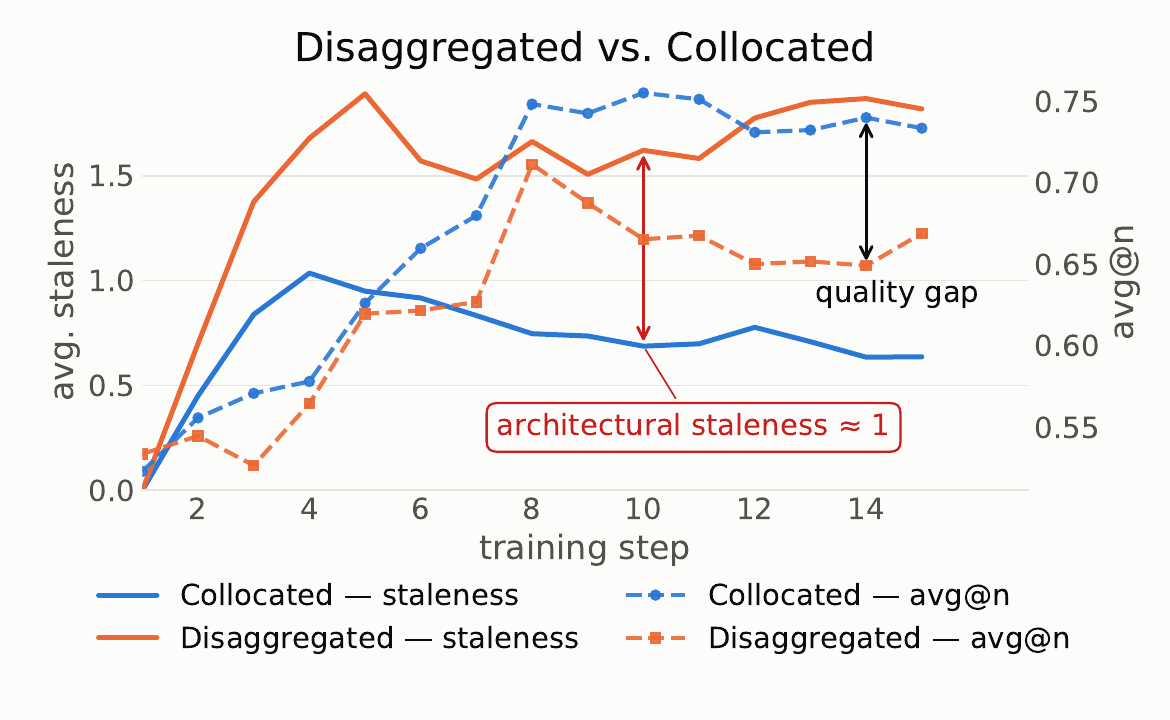}
    \end{minipage}\\[-3.2pt]
    \begin{minipage}[t]{0.38\linewidth}
      \centering\small (a)
    \end{minipage}%
    \hspace{0.02\linewidth}%
    \begin{minipage}[t]{0.45\linewidth}
      \centering\small (b)
    \end{minipage}
  \par\vspace{-0.12in}
  \caption{\label{fig:architectural-staleness} Architectural staleness under partial rollout.
  (a) Colocated and disaggregated
  execution with batch-level streaming, where $i$ denotes the $i$-th version of parameters, 
  assuming the durations of the two stages are balanced. (b) Average staleness and avg@$n$ of VeRL
      (colocated) and slime (disaggregated).}
  \vspace{-0.15in}
\end{figure*}

\section{Background}
\label{sec:background}

We briefly introduce the background of this work from the
perspectives of RL trajectories and system architectures.

\subsection{Trajectory in RL Systems}
\label{sec:background-trajectory}

A \emph{trajectory} records one rollout from an input prompt,
including generated tokens and any intermediate interactions.
We use \emph{sample} to refer to the prompt together with its
associated trajectories. For example, GRPO-style algorithms
roll out the same prompt $n$ times~\cite{shao2024deepseekmath},
producing $n$ trajectories. Beyond the generated token IDs and
interaction history, trajectory payloads also include rewards
and per-token auxiliary data such as routing decisions~\cite{r3}
and top-$p$ sampling masks~\cite{mai_thinking_1}, which typically
occupy tens to thousands of times as many bytes as the token IDs.
Trajectories may also carry multimodal inputs such as images
and videos, which can account for an even larger share of
the payload in some tasks~\cite{mmrl,kimi2026k25}.
When the prompt contains multimodal inputs, these trajectories
share the same inputs.

We use \emph{trajectory transfer} to denote the end-to-end process
of making completed rollout trajectories available to training,
including admission, storage, and data preparation and delivery
to training workers. Trajectory completion and training admission
are distinct: completion means that a trajectory has finished
rollout, whereas admission involves not only determining whether
it is accepted under the RL algorithm, but also whether it is
selected to form the next training batch. This distinction is
particularly relevant in streaming rollout, where trajectories
progress individually rather than as a fixed rollout batch
~\cite{zhong2025streamrl,fu2025areal}.
The transfer path involves a batching hierarchy of trajectories
already present in RL training
~\cite{schulman2017proximal,sheng2025hybridflow,yu2025dapo}.
A \emph{train batch} is the batch whose training completion
triggers parameter synchronization in disaggregated systems or
the stage switch in colocated systems. A \emph{mini-batch} is
the unit whose completion triggers a parameter update on the
training side. A \emph{micro-batch} is the unit of one
forward--backward pass.

\begin{figure*}[t]
  \centering
    \begin{minipage}[c]{0.30\linewidth}
      \centering
      \includegraphics[width=\linewidth]{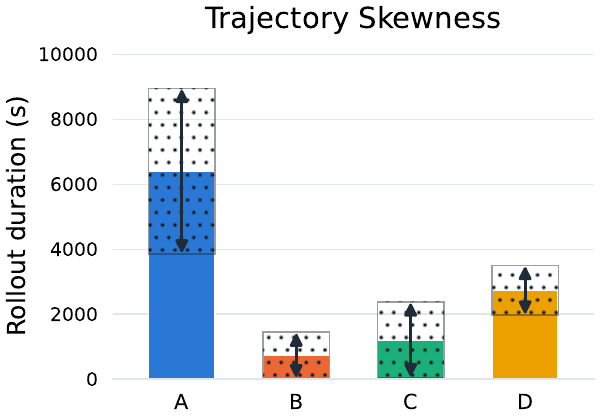}
    \end{minipage}%
    \hspace{0.02\linewidth}%
    \begin{minipage}[c]{0.30\linewidth}
      \centering
      \includegraphics[width=\linewidth]{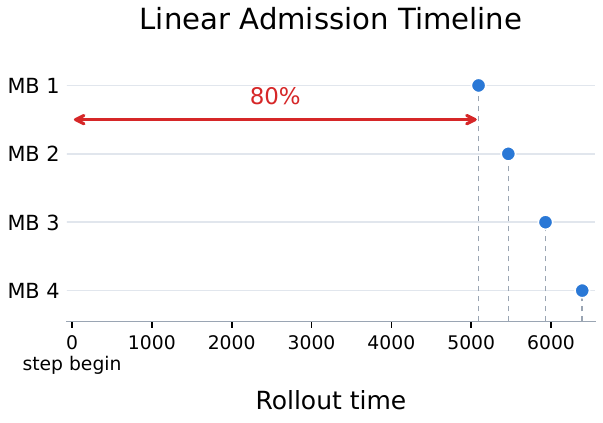}
    \end{minipage}%
    \hspace{0.02\linewidth}%
    \begin{minipage}[c]{0.34\linewidth}
      \centering
      \includegraphics[width=\linewidth]{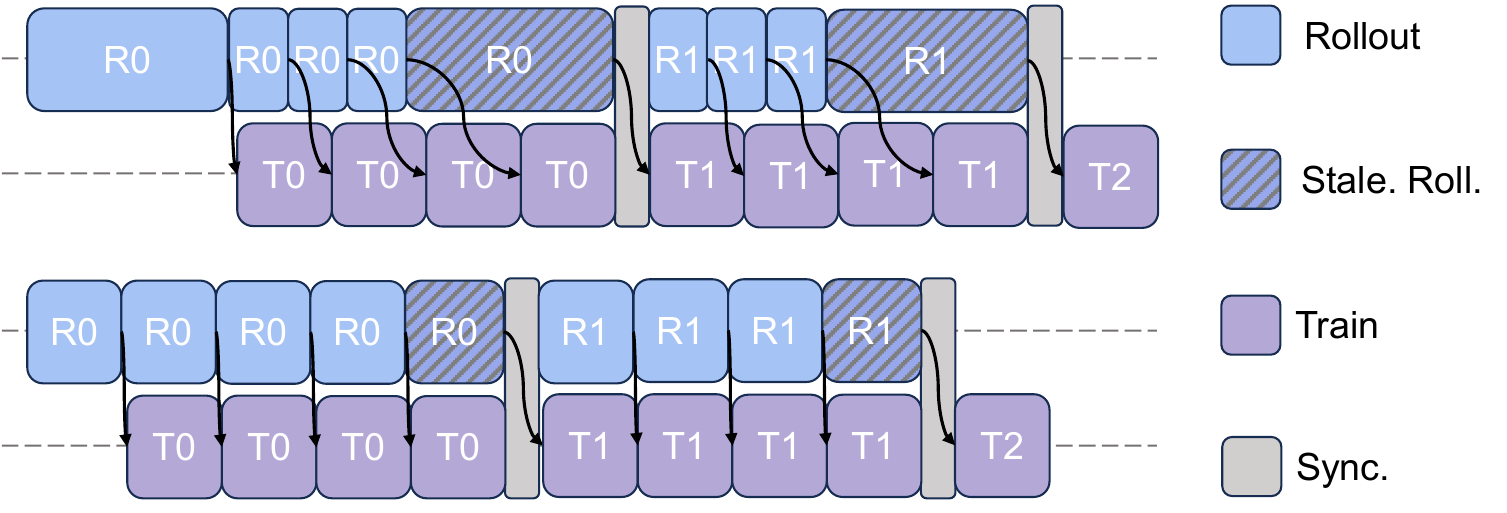}
    \end{minipage}\\[1pt]
    \makebox[0.30\linewidth]{\small (a)}%
    \hspace{0.02\linewidth}%
    \makebox[0.30\linewidth]{\small (b)}%
    \hspace{0.02\linewidth}%
    \makebox[0.34\linewidth]{\small (c)}
  \par\vspace{-0.12in}
  \caption{\label{fig:admission-skewness} Rollout skew, delayed admission and larger architectural staleness.
    (a) Per-source rollout durations: bars show means;
    arrows show $\pm$ one standard deviation.
    (b) Mini-batch admission times under linear cumulative source targets.
    (c) Delayed (top) versus earlier (bottom) admission at the same mini-batch granularity.}
  \vspace{-0.15in}
\end{figure*}

\subsection{Architectural Staleness}
\label{sec:motivation-staleness}

RL systems can be classified by where rollout and training are
placed: disaggregated systems dedicate a separate resource pool to
each stage, allowing the stages to be pipelined
~\cite{hu2024openrlhf,zhong2025streamrl,thudm2025slime},
while colocated systems time-share one pool of devices between
rollout and training and switch stages at their boundaries
~\cite{yao2023deepspeed,sheng2025hybridflow}.
Regardless of placement, waiting for all scheduled trajectories to
finish can expose training to long-tail stalls, whether execution is
synchronous or one-step asynchronous
~\cite{huang2023cleanba,noukhovitch2025asynchronous}.
Partial rollout, an off-policy approach to reducing such waits, pauses
unfinished trajectories and resumes generation with updated parameters
in later training steps. This technique applies not only to
disaggregated systems but also to colocated systems~\cite{kimi2025k15}.

The \emph{staleness} measures the gap in parameter versions between
data generation and training. It is defined per token: a token
generated with parameter version $v$ and trained at version $v'$ has
staleness $v' - v$;
the staleness of a trajectory usually refers to that of its oldest
token. Stale
trajectories are off-policy for the current model, and excessive staleness can degrade training
quality. 
Existing asynchronous RL systems incorporate staleness-control
mechanisms to limit this version gap~\cite{fu2025areal,li2026staleflow}.
However, disaggregated pipelining
~\cite{luo2025deepcoder,noukhovitch2025asynchronous}
introduces \emph{architectural staleness} beyond that caused by
cross-step generation: rollout continues during $T_i$, but
the resulting data is consumed by $T_{i+1}$, after the parameters
have advanced by one version
(Figure~\ref{fig:architectural-staleness}(a), where shading
highlights rollout intervals whose generated tokens are stale
due to pipelining).
Figure~\ref{fig:architectural-staleness}(b) confirms this
in practice: under the same configuration and staleness-control 
algorithm, the disaggregated system has an average staleness 
about one step higher than that of the colocated system. Its avg@$n$
--- the mean reward over the $n$ rollouts of a prompt --- 
falls behind accordingly. These observations motivate
reducing architectural staleness in the rollout-to-training path.

\section{Motivation}
\label{sec:motivation}

We examine two challenges along the rollout-to-training path:
enabling early admission while preserving batch-wide training
requirements (\S\ref{sec:motivation-admission}), and aligning
trajectory data access with the spatial and temporal needs of
distributed training (\S\ref{sec:motivation-storage}).

\subsection{Admission under Train-Batch Constraints}
\label{sec:motivation-admission}

RL training can impose batch-wide requirements that cannot be
checked on each trajectory independently. Mixed RL uses target
mixing ratios across data sources~\cite{glm2026glm5,datamix},
while partial-rollout completion quotas require a minimum
completion ratio for each rollout step to mitigate long-tail
starvation. These requirements govern either the composition
of a train batch or the rollout progress required for admission.
Enforcing these requirements by assembling a complete train
batch before releasing any trajectories creates a
\emph{train-batch admission barrier}: training waits until all
admission conditions hold, even when some completed trajectories
have already passed their individual acceptance checks.

Existing trajectory streaming systems
~\cite{zhong2025streamrl,han2025asyncflow} allow training to
consume smaller groups of trajectories as rollout progresses,
but do not address admission under these batch-wide requirements.
When rollout completion times are skewed across data sources
(Figure~\ref{fig:admission-skewness}(a)), admitting trajectories
as soon as they complete can violate the final batch composition.
For example, consider a train
batch of 100 trajectories requiring 50 from each of two sources,
A and B, divided into four mini-batches of 25. If A completes
faster, admitting 75 trajectories from A in the first three
mini-batches leaves only 25 slots for B. Even filling the last
mini-batch entirely with B yields a 75:25 composition rather
than the required 50:50. Once these trajectories enter training,
later admissions cannot repair the earlier allocation.
Retaining the train-batch admission barrier to enforce these
requirements, however, blocks early consumption and preserves
the batch-level pipeline illustrated in
Figure~\ref{fig:architectural-staleness}(a).
The associated architectural staleness degrades training quality
(\S\ref{sec:motivation-staleness}).

A natural strawman is to scale each source's cumulative quota
linearly with mini-batch admission progress. Under such skew,
however, requiring every source to reach a proportional target
at each admission boundary can delay even the first mini-batch.
With four mini-batches per train batch,
Figure~\ref{fig:admission-skewness}(b) shows that nearly 80\%
of the interval from rollout start to final admission elapses
before the first mini-batch is admitted.
Figure~\ref{fig:admission-skewness}(c) compares schedules at the
same mini-batch granularity: delayed admission (top) postpones
training completion and parameter synchronization, increasing
the shaded rollout fraction and thus architectural staleness
relative to earlier admission (bottom).
Thus, finer admission granularity alone does not ensure early
streaming. The challenge is to enable
early mini-batch admission under uneven rollout progress by
overcoming the train-batch admission barrier while preserving
the final train-batch requirements. \sysname addresses this
challenge through \emph{Constraint-Preserving Mini-Batch
Admission} (\S\ref{sec:design-admission}).

\subsection{Trajectory Data Access}
\label{sec:motivation-storage}

\begin{figure}[t]
  \centering
    \includegraphics[width=\linewidth]{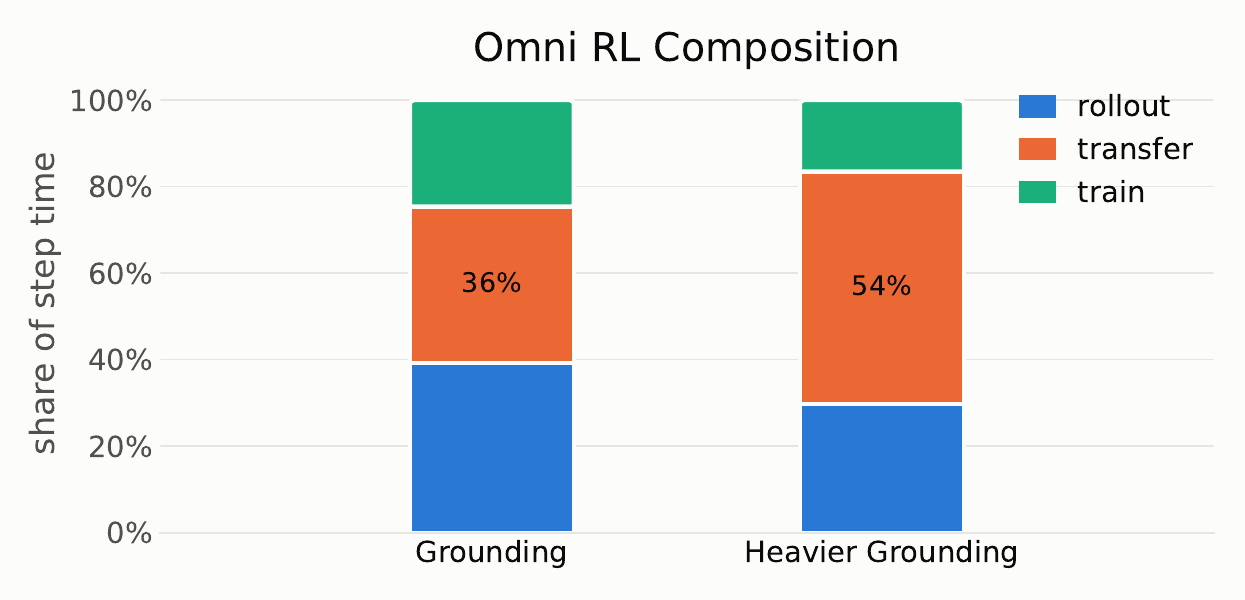}
  \par\vspace{-0.15in}
  \caption{\label{fig:omni-transfer-share} RL step-time breakdown on two
    Omni RL workloads using VeRL with sample-level transfer.}
  \vspace{-0.15in}
\end{figure}

Even after trajectories are admitted, preparing and delivering
their large payloads can delay training. In pre-training,
inputs for the next training step can be prepared from an
existing dataset while the current step
executes~\cite{zhao2026megascale}. RL instead consumes
trajectories generated online by rollout, making it difficult
to hide data preparation entirely in the background. Data
preparation and delivery thus remain on the critical path
between rollout and training.
On two colocated Omni RL workloads,
rollout-to-training transfer accounts for 36\% and 54\% of a
training step, respectively, exceeding the training computation
time in both cases (Figure~\ref{fig:omni-transfer-share}).

\parabf{Storage Granularity and Spatial Demand.}
Storage organization determines which data can be accessed
independently, while the training topology determines which
data each worker needs. Existing sample-level storage
layouts~\cite{han2025asyncflow,mooncake2026miles} organize data
around sample boundaries rather than these consumer-specific
requirements. 
Large-scale training combines data parallelism
(DP), context parallelism (CP), and tensor parallelism (TP).
TP requires all ranks in a TP group to process bit-wise identical
micro-batches, whereas CP allows each rank to process only a
slice of a long sequence. A sample-level layout therefore either
repeatedly transfers the same data to multiple ranks or transfers
whole samples when a rank needs only a subset of their data.
Alternatively, avoiding these redundant transfers requires
fetching once and broadcasting (TP) or scattering (CP) locally,
which removes the duplication but inserts an extra synchronization
hop whose latency lands on the critical path, slowing down the
training process.
Omni RL further exposes this mismatch because multimodal
encoders can have assignments distinct from the backbone's.
Their encoding cost is determined by the multimodal inputs
and requires separate DP balancing~\cite{zheng2025orchmllm};
their TP and CP degrees can also differ from the
backbone's~\cite{shoeybi2019megatron,feng2024optimus}.
Some systems even place encoders in a separate
cluster~\cite{zhang2024disttrain}.
When multimodal inputs are fetched according to the backbone's
sample assignment, they must then be redistributed to the ranks
assigned to encode them, introducing additional data transfer.
Data access must therefore account for the encoder and backbone
topologies separately, rather than using the backbone's assignment
for all trajectory data.

\parabf{Preparation Granularity and Temporal Demand.}
An admitted mini-batch does not require all of its data at once:
training consumes it through a sequence of smaller micro-batches.
Unlike in pre-training, where inputs can be prepared a step
ahead, retrieving large payloads from distributed storage and
packing them into training inputs remain time-consuming in RL,
even without redundant data movement.
Existing RL systems such as VeRL~\cite{sheng2025hybridflow}
and slime~\cite{thudm2025slime} support micro-batch execution,
but their batch-oriented training paths leave payload retrieval
and input packing serialized with training computation on the critical
path, further prolonging training time.
The underlying issue is the lack of on-demand data preparation
aligned with the temporal needs of individual micro-batches.

\smallskip\noindent The challenge of trajectory data access is therefore to align trajectory data access with
both where data are consumed and when they are needed, avoiding
redundant movement and unnecessary waiting. \sysname addresses
this challenge through \emph{On-Demand Trajectory Data Access}.

\section{System Overview}
\label{sec:overview}

\begin{figure}[t]
  \centering
    \includegraphics[width=0.88\linewidth]{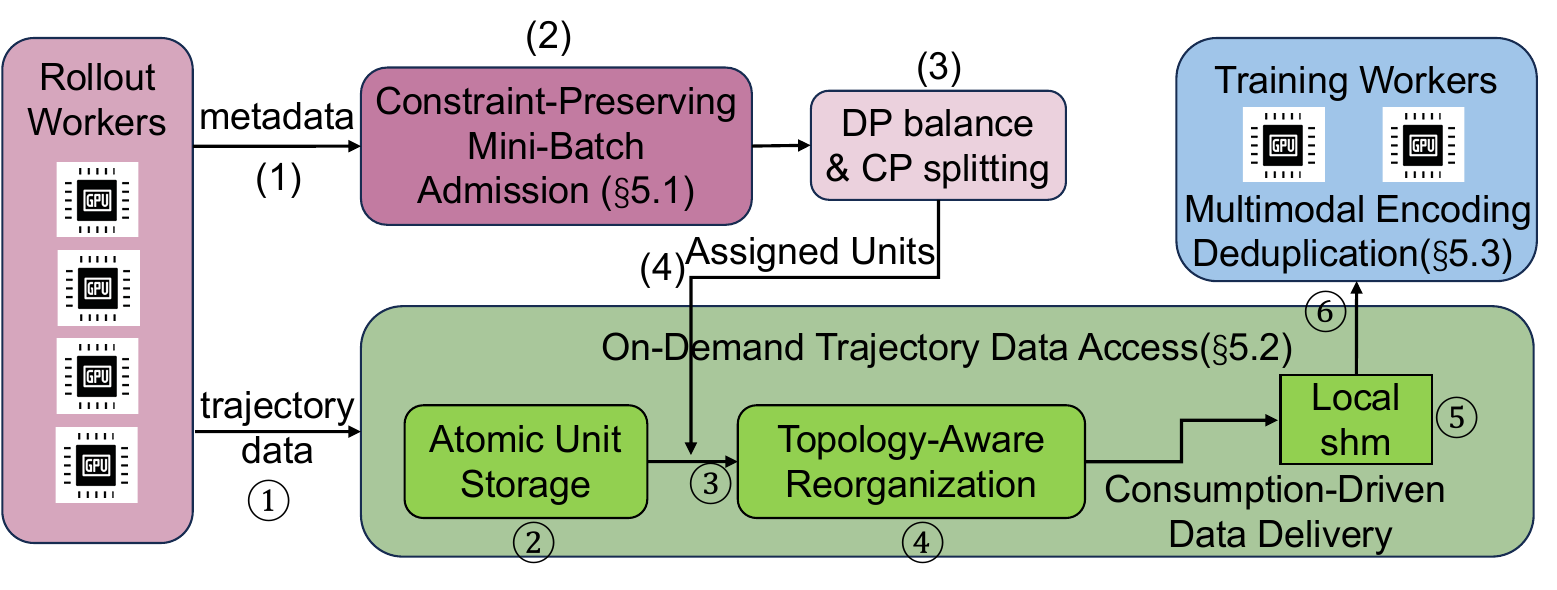}
  \par\vspace{-0.15in}
  \caption{\label{fig:overview} Overview of \sysname.}
  \vspace{-0.15in}
\end{figure}

Figure~\ref{fig:overview} shows the end-to-end data flow of
\sysname, from completed rollout trajectories to their
admission and consumption by training workers.

When a trajectory completes, rollout workers report its metadata
to the control plane (1) and write its payload to distributed
storage \textcircled{1}. Atomic-unit storage \textcircled{2}
records the payload as independently addressable units organized
through a hierarchical index, allowing subsequent access according
to training assignments.

Using the reported metadata, the control plane admits a
mini-batch when its progressive readiness conditions and
feasibility guards hold (2) (\S\ref{sec:design-admission}).
It then determines DP balancing and CP splitting according to
each consumer's topology (3), and dispatches the resulting
assignments (4) to TP-group-local reorganization processes.
Following these assignments, the processes retrieve the required
atomic units, using sliced reads for assigned sequence ranges
\textcircled{3}, and pack them into training inputs \textcircled{4},
publishing them incrementally in local shared memory in each
consumer's micro-batch order \textcircled{5}.
Training workers prefetch the prepared inputs into pinned CPU
memory and transfer them to the GPU \textcircled{6}, overlapping
preparation and staging of subsequent micro-batches with ongoing
computation (\S\ref{sec:design-storage}).
For multimodal workloads, training workers reuse each unique
input's encoding across the admitted mini-batch and aggregate
consumer gradients for encoder backward, while the encoder and
backbone retain independent micro-batch partitions
(\S\ref{sec:design-efficiency}).

\section{Design}
\label{sec:design}

The key is to separate logical training boundaries from physical
execution granularity. Train-batch requirements need not force
train-batch admission (\S\ref{sec:design-admission}); admitting a
mini-batch need not require materializing all of its data at once
(\S\ref{sec:design-storage}); and micro-batch execution need not
repeat the encoding of shared inputs (\S\ref{sec:design-efficiency}).

\subsection{Constraint-Preserving Mini-Batch Admission}
\label{sec:design-admission}

\sysname forms mini-batches for admission from completed rollout
trajectories by combining progressive rollout-side readiness
conditions with feasibility guards on candidate batch composition.
We first express the original train-batch admission gate as a
conjunction of predicates:
\begin{equation}
  \textsc{BatchReady} = \textstyle\bigwedge_i P_i,
\end{equation}
where each predicate $P_i$ captures one requirement on the train batch,
such as reaching the train-batch size, meeting a partial-rollout
completion quota, or maintaining a target data-source ratio
~\cite{datamix}. For example, ROLL assembles a training batch only
after every domain meets its quota.

For quota-based requirements, we first define rollout-side readiness conditions:
\begin{equation}
  P_i:\; s_i \ge \tau_i,
\end{equation}
where $s_i$ is the rollout-side count for quota $i$, and $\tau_i$
is its final target. For example, if a
data source should contribute at least a fraction $r_i$ of a train
batch of size $\tau_0$, the corresponding readiness condition is
$s_i \ge \tau_i$, where
$\tau_i=r_i\tau_0$; analogous quotas apply to the other sources. Let
$s_{i,0}$ denote the count at the beginning of the current train
batch. For the $t$-th mini-batch, we set a progressive target between
$s_{i,0}$ and $\tau_i$ using predicate-specific interpolation:
\begin{equation}
  \tau_{i,t} = s_{i,0} + \alpha_{i,t}(\tau_i - s_{i,0}),
  \quad t = 1, \dots, n,
\end{equation}
where $0 \le \alpha_{i,1} \le \cdots \le \alpha_{i,n}=1$.
The coefficients specify a configured schedule for each predicate.
For equal-sized mini-batches, the train-batch-size target grows
linearly. Source targets instead use relaxed early coefficients
and reach the original quotas at the final admission. This
accommodates uneven rollout progress without requiring every
readiness target to grow in proportion to the train-batch size.
Before admitting the $t$-th mini-batch, the rollout-side counts
must satisfy $P_{i,t}: s_i \ge \tau_{i,t}$, allowing earlier
mini-batches to proceed before all final targets are reached.
Figure~\ref{fig:intro} illustrates one such schedule, with
cumulative source-readiness targets $(0,0,2,4)$ instead of
linear targets $(1,2,3,4)$. Both schedules reach the same final
source quotas, but the former permits earlier admission under
the illustrated rollout skew.

However, progressive readiness alone does not ensure
final train-batch feasibility. Consider the two-source example
from \S\ref{sec:motivation-admission}: a train batch requires
50 trajectories from each source, admitted through four
mini-batches of 25. Suppose each source's readiness target is
10 at the third admission and reaches 50 at the final admission.
The current readiness conditions can hold even when the first
three admitted mini-batches contain 65 trajectories from A and
10 from B. Only 25 slots then remain for the 40 additional
trajectories required from B, making the final quotas unattainable.

We therefore append \emph{feasibility guards} to the per-mini-batch
predicates.
Once all progressive rollout-side readiness predicates $P_{i,t}$
hold, an application-specific selection policy proposes a
candidate mini-batch. We then apply the guards to check whether
the candidate can be admitted without making the final train-batch quotas
unattainable. Let $b_{i,t}$ denote the cumulative count of class
$i$ in the train batch if the candidate mini-batch is admitted,
in contrast to the rollout-side count $s_i$. Let $b_{0,t}$
denote the total admitted count including this candidate.
We partition batch-composition quotas into groups
$G_1, \dots, G_m$ whose target classes compete for the same batch
slots. For example, data-source quotas form one group, while
partial-rollout completion quotas remain rollout-side readiness
conditions rather than batch-capacity guards. For the candidate's
prospective status $B_t = \{b_{i,t}\}$, we check one guard for
each group:
\begin{equation}
  Q_k(B_t)\colon\;
  \textstyle\sum_{i \in G_k} \bigl(\tau_i - b_{i,t}\bigr)^{+}
  \;\le\; \tau_0 - b_{0,t},
  \label{eq:admission-guard}
\end{equation}
where $(x)^{+} = \max(x, 0)$. The right-hand side is the remaining
capacity of the train batch, while the left-hand side is the
outstanding quota deficit for the group. The guards prevent an
irrevocable mini-batch decision from making the final train batch
infeasible. The guard rejects any candidate that would produce
the composition above. The admission predicate is therefore
\begin{equation}
  \textsc{BatchReady}_t =
  \textstyle\bigwedge_i P_{i,t} \;\wedge\;
  \textstyle\bigwedge_k Q_k(B_t).
  \label{eq:admission-predicate}
\end{equation}
The candidate for mini-batch $t$ is admitted only when
$\textsc{BatchReady}_t$ holds.
Individual mini-batches may therefore deviate from the final
train-batch quotas, while the full train batch meets all original
requirements.

\subsection{On-Demand Trajectory Data Access}
\label{sec:design-storage}

\begin{figure}[t]
  \centering
    \includegraphics[width=0.88\linewidth]{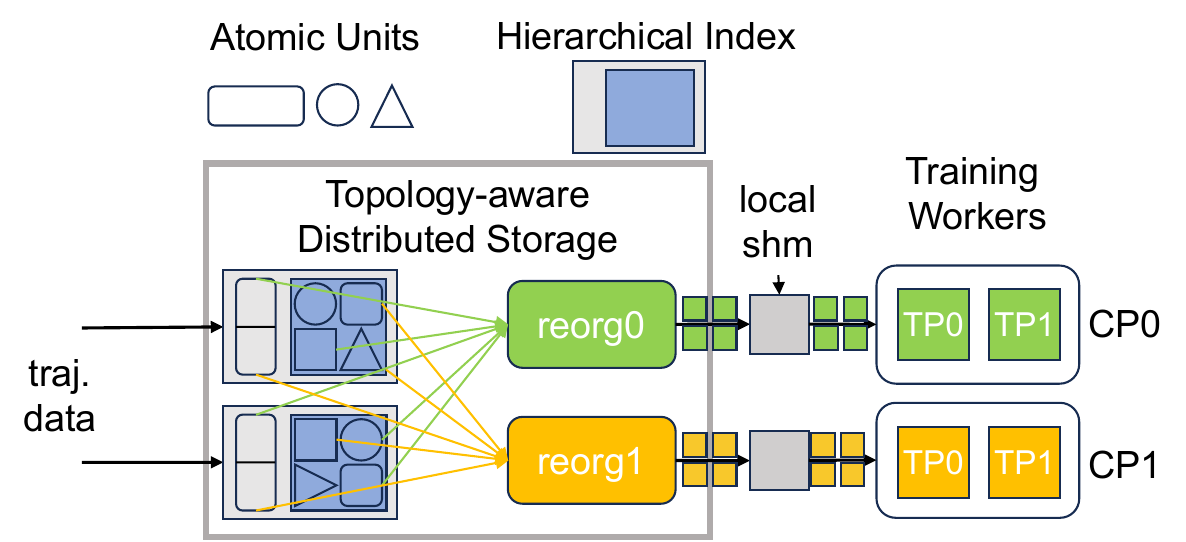}
  \par\vspace{-0.07in}
  \caption{\label{fig:on-demand} End-to-end workflow of On-Demand Trajectory Data Access.}
  \vspace{-0.1in}
\end{figure}

\sysname aligns trajectory data access with the spatial and
temporal needs of training (Figure~\ref{fig:on-demand}).
Atomic-unit storage and topology-aware
reorganization determine which data are prepared for each
consumer. Together, they form the \emph{Topology-Aware Distributed
Storage} shown in Figure~\ref{fig:on-demand}.
Consumption-driven delivery coordinates data preparation with
training progress within an admitted mini-batch.

\parabf{Atomic-Unit Storage.}
When the rollout stage writes into distributed storage, \sysname
stores data as independently addressable atomic units that match
the DP-balancing granularity specified by each training topology.
In GRPO, for example, a complete trajectory is one atomic unit: once
admitted for training, it is batched and processed independently of
the other trajectories generated from the same prompt, so the sample
as a whole is not atomic.
The storage representation does not fix CP slices at write time:
CP splitting depends on the training batch and is determined
after admission. For multimodal data, an atomic unit is an image or
a group of consecutive video frames, as defined by the model architecture.
A complete sample is then recorded with a hierarchical index: a sample
maps to the storage keys of its trajectories, and each trajectory in
turn maps to the storage keys of the other atomic units it references,
mainly multimodal ones. The hierarchical index allows each atomic
unit to be addressed independently, as illustrated in Figure~\ref{fig:on-demand}.
This storage representation is defined at the logical storage
layer and is independent of the physical backend, allowing the
same atomic-unit abstraction to operate over Mooncake, the Ray
object store, or other backends.

\parabf{Topology-Aware Reorganization.}
When a mini-batch is admitted, the RL control plane determines data
assignments according to each consumer's topology, performing
DP balancing~\cite{verlbalancing,jiang2024dynapipe,zheng2025orchmllm} and CP
splitting~\cite{fang2024usp} based on metadata such as trajectory
lengths and the computation costs of other atomic units, helping to
reduce memory pressure.
Each assignment is dispatched to a \emph{reorganization process}
on the local node of the corresponding TP group.
Each process serves one CP partition and is shared by the TP
ranks processing that partition. This organization addresses
two complementary requirements:
\begin{itemize}
  \item A separate process for each \textbf{CP} partition can retrieve only
  the sequence slices assigned to it within the admitted mini-batch.
  To support these reads uniformly in the logical storage layer,
  \sysname adds a sequence-length field to atomic storage units
  where applicable
  and extends the read interface with an optional sequence-offset
  parameter. The logical storage layer coordinates with the
  physical backend to perform the corresponding slice reads,
  supporting both memory-based backends
  ~\cite{transferqueue2026,qin2025mooncake} and disk-based backends.

  \item Sharing one process across a \textbf{TP} group avoids repeated retrieval
  of the same trajectory data and centralizes input packing for
  the group's forward computation. Since TP ranks are typically
  colocated on a physical node for efficient communication,
  node-local shared memory can distribute the prepared inputs
  without an extra fetch-then-broadcast step among the ranks.
\end{itemize}

\smallskip\noindent As shown in Figure~\ref{fig:on-demand}, the
process follows the hierarchical indices to locate the required
atomic units. It then uses the training framework's packing
function to prepare the group's inputs, including concatenating
trajectories, applying padding where
needed, and constructing attention masks or sequence-boundary
metadata for packed attention~\cite{jiang2024dynapipe,nvidianemopacking}.
The resulting inputs are published in local shared memory for
the group's training workers.
In Omni RL, multimodal encoders have distinct topologies, so
each encoder type has its own storage configuration and
reorganization processes, applying the same retrieval, packing,
and publication flow separately from the backbone.

\parabf{Consumption-Driven Data Delivery.}
The assignment is determined over a mini-batch, but its payload
need not be materialized all at once. The reorganization process
executes reads and packing incrementally in each consumer's
micro-batch consumption order, notifying the training workers as prepared
inputs become available in shared memory. Each training worker
uses a background thread to prefetch these inputs into pinned
CPU memory. The training loop then asynchronously copies each
prefetched micro-batch to the GPU for computation, instead of
waiting for the entire mini-batch to become ready.
Figure~\ref{fig:intro} shows how the multimodal and
backbone reorganization processes incrementally supply their
respective computation stages. This coordination allows both
reorganization and pinned-memory staging for later inputs to
overlap with GPU computation on earlier inputs. This overlap reduces
data-preparation delays on the training critical path without changing
mini-batch admission or optimizer-update boundaries.

\subsection{Multimodal Encoding Deduplication}
\label{sec:design-efficiency}

GRPO-style RL produces multiple trajectories that share the same
multimodal inputs. Using the storage keys recorded in the hierarchical
index (\S\ref{sec:design-storage}), \sysname identifies shared
multimodal inputs within each mini-batch. However, sharing stored input
data does not itself eliminate repeated encoding during training.
The training-side challenge is a mismatch in execution scope:
encoding reuse spans an entire mini-batch, while the backbone
consumes the shared encodings through separate micro-batch forward
and backward passes. Without additional handling, the first backward
pass through a shared encoding releases the saved encoder activations
needed by later micro-batches. Keeping the encoded tensor available
is therefore insufficient to reuse its encoder computation graph
across all consumers.

\begin{lstlisting}[
  language=Python,
  basicstyle=\ttfamily\footnotesize,
  keywordstyle=\bfseries,
  commentstyle=\bfseries,
  numbers=left,
  numberstyle=\tiny,
  numbersep=5pt,
  xleftmargin=1.5em,
  columns=fullflexible,
  keepspaces=true,
  breaklines=true,
  showstringspaces=false,
  frame=tb,
  float=t,
  captionpos=b,
  caption={Mini-batch-wide multimodal encoding reuse.},
  label={lst:encoding-dedup}
]
def train_minibatch(batch, plan):
    optimizer.zero_grad()
    encoded = []
    # Encode unique inputs; retain graphs
    for micro in encoder_micro_batches(batch, plan):
        outputs = encoder(micro.inputs)
        encoded.append((micro.ids, outputs))
    # Topology-aware routing and caching
    cache, refs, route = route_detached(
        encoded, batch, plan)
    cache_grad = zeros_like(cache)
    # Reuse encodings; accumulate gradients
    for micro in backbone_micro_batches(batch):
        ids = refs[micro.id]
        features = cache[ids].detach().requires_grad_()
        result = backbone(micro, features)
        normalized_loss(result).backward()
        cache_grad.index_add_(0, ids, features.grad)
    # Return gradients; deferred encoder backward
    owned_grad = tp_owned_slices(cache_grad, route)
    encoder_grad = reverse_route_sum(owned_grad, route)
    for ids, outputs in encoded:
        backward(outputs, encoder_grad[ids])
    finalize_model_grads()
    optimizer.step()
\end{lstlisting}

\sysname enables mini-batch-wide training with a single encoding
of each unique input, preserving the gradient contributions of
all consuming trajectories.
First, \sysname stores each unique multimodal input as a single atomic
unit in distributed storage, with all references to that input sharing
the same storage key. The control plane also treats each unique input as
a single unit for encoder-side DP balancing, assigning it to only one
group of ranks for encoding.
\sysname then separates the lifetime of the encoder graph from the
backbone's micro-batch execution. It encodes each unique multimodal
input once per mini-batch and retains the corresponding encoder
graph. Since multimodal inputs and backbone data are independently
DP-balanced, the encoding and consuming ranks need not coincide.
Using both sides' topologies and assignments, \sysname maps each
unique encoding's source to all consuming destinations, including
trajectory repeats, TP replicas, and copies induced by CP boundaries.
It distributes detached encoding values along this mapping and
caches them at the consuming ranks for reuse within the mini-batch.

Each backbone micro-batch gathers its required encodings from the
local cache as detached inputs that require gradients. The backbone's
backward passes stop at this boundary without traversing the encoder graph.
The resulting gradients are accumulated locally by encoding identity,
combining contributions from repeated references within and across
micro-batches.
After all backbone micro-batches finish, \sysname returns and
aggregates the accumulated gradients according to the encoding distribution
mapping. Each TP rank contributes only its assigned gradient slice,
avoiding duplicate accumulation across TP replicas. The encoding
ranks then backpropagate the aggregated gradients through the
retained encoder graphs before the optimizer update.
Figure~\ref{fig:intro} places the encoder's forward pass before
the backbone micro-batch passes and its backward pass after them,
reflecting this deferred encoder backward. Caching encodings
and accumulating their gradients locally avoids repeating
encoder--backbone transfers for each micro-batch.
Regarding the additional GPU memory required to cache encodings
throughout the mini-batch, deduplication reduces the number of cached
encodings, while releasing micro-batch-local copies as gradients
accumulate before encoder backward alleviates peak memory pressure.

The encoder partitions the mini-batch's unique multimodal inputs
into its own micro-batches, independently of the backbone's
micro-batch partitioning. Consumption-driven delivery
(\S\ref{sec:design-storage}) overlaps preparation of later encoder
micro-batches with computation on earlier ones. The encoder
graphs remain available until the gradients from all backbone
micro-batches have been accumulated and returned for encoder backpropagation.
Thus, both modules retain their own execution granularity while
encoding reuse spans the entire mini-batch.
Listing~\ref{lst:encoding-dedup} summarizes the forward and backward
flow for one mini-batch.

\section{Implementation}
\label{sec:implementation}

\sysname is built on VeRL, with SGLang as the inference engine
and Megatron as the training engine. We extend VeRL with partial rollout
~\cite{kimi2025k15} and slime-style disaggregated execution.
Its topology-aware distributed storage builds on
TransferQueue~\cite{transferqueue2026}, an
open-source library for data storage in post-training workflows
that disaggregates the control plane and data plane in RL trajectory transfer.

\parabf{\texttt{TQClient} Extensions.}
We override the write interfaces of TransferQueue's \texttt{TQClient}
to support both whole-trajectory writes and updates to individual
fields. Each field's topology identifier selects the corresponding
splitting function, which decomposes its data into atomic units.
The client stores these units and returns the index entries for
the updated fields.
On the training side, initializing \texttt{TQClient} with a
\texttt{TopologyConfig} connects workers to a shared node-local Ray
actor. After accepting admission and assignment results, the actor
retrieves the required atomic units from \texttt{TQClient}.
The actor then packs the retrieved data and materializes the
prepared inputs in Ray's Plasma object store.
Local workers share the materialized arrays, while
Ray's inter-node object transfer also supports TP groups spanning
multiple nodes. For NumPy-compatible CPU tensors, we expose NumPy
views and reconstruct tensor views at the receivers, allowing local
workers to access the shared arrays without additional copies.
Multimodal data use a
separate \texttt{TQClient} configured for the encoder topology,
following the same storage and materialization path.
By keeping this refactoring entirely within TransferQueue, we make
the design of \sysname easy to adapt to other RL frameworks.

\parabf{Sliced Reads.}
At the logical layer, \texttt{TQClient}'s read interface accepts
optional \texttt{field\_ranges} specifying per-field, per-sample
sequence ranges, which the storage manager converts to byte
offsets and slice shapes when required by the backend.
At the physical layer, TransferQueue's in-memory storage units
slice tensors before transmission, Mooncake uses
\texttt{get\_into\_ranges} to fetch the requested bytes, and the
disk backend reads the corresponding ranges from shared-filesystem
shards with \texttt{preadv}.

\parabf{Micro-Batch Delivery.}
As each micro-batch is materialized, the reorganization actor publishes
its Ray object handle in a thread-safe dictionary indexed by
micro-batch. Training workers call
\texttt{await tq\_client.get\_next\_micro()} to retrieve the next
handle, obtaining the materialized
layout without waiting for the entire mini-batch. After the
first-layer kernel of the current micro-batch is launched, CUDA
events coordinate staging of the next micro-batch into pinned host
memory.

\setcounter{topnumber}{2}
\setcounter{dbltopnumber}{2}

\section{Evaluation}
\label{sec:eval}

We evaluate \sysname's end-to-end performance
(\S\ref{sec:eval-e2e}) and RL quality
(\S\ref{sec:eval-quality}), and use ablation studies
(\S\ref{sec:eval-ablation}) to isolate the contributions
of individual designs.

\subsection{Experimental Setup}
\label{sec:eval-setup}

\parabf{Tasks and Datasets.} We evaluate \sysname on two RL
tasks. The first is a \textbf{CodeAgent} task, for which we train
on a mixture of well-known software engineering datasets:
SWE-smith~\cite{yang2025swesmith},
Multi-SWE-bench~\cite{zan2025multiswebench},
Terminal-Lego~\cite{yang2026terminallego}, and
Scale-SWE~\cite{zhao2026scaleswe}. The second is \textbf{Video Grounding},
representing multimodal tasks. We use two in-house datasets,
Grounding A and Grounding B, that differ in their fraction of
video data to cover different multimodal payloads. The maximum
context length is 256K for CodeAgent and 64K for Video Grounding.

\parabf{Models.} We train three large models for large-scale RL
experiments: Qwen3-VL-235B
~\cite{qwen3vl}, MiMo-V2.5-309B~\cite{mimov2}, and DeepSeek-v3.2-671B
~\cite{deepseekv32}. The first two understand multimodal inputs and
are trained on both tasks, whereas DeepSeek-v3.2 is trained
only on the CodeAgent task. All models use FP8 quantization for both
training and inference.

\parabf{Baselines.} Our baselines are
VeRL~\cite{sheng2025hybridflow} for colocated execution and
slime~\cite{thudm2025slime} for disaggregated execution.
We align their staleness control algorithms to isolate 
the architectural contribution to staleness.
We also use StreamRL~\cite{zhong2025streamrl} and
AsyncFlow~\cite{han2025asyncflow} as streaming baselines without
batch-wide admission constraints.
For distributed storage and data movement, we compare three
configurations within the same training framework:
\begin{itemize}
  \item \textbf{TransferQueue} (\texttt{TQ})
  ~\cite{han2025asyncflow,transferqueue2026}: an open-source Python library
  with sample-level storage and no topology-aware reorganization.
  Each rank reads the backbone data for its assigned trajectories.
  Only backbone ranks with both TP rank 0 and CP rank 0 read multimodal data,
  which are then redistributed through an all-to-all collective over
  the global communication group according to the multimodal scheduling plan.
  \item \textbf{Miles} (\texttt{Miles})
  ~\cite{mooncake2026miles}: uses sample-level storage, like TransferQueue,
  with a Mooncake KV backend and RDMA transport for efficient data access.
  \item \textbf{MegaScale-Data} (\texttt{MegaData}$^{*}$)
  ~\cite{zhao2026megascale}: a topology-aware data preprocessing
    system designed for pre-training and adapted here to RL
    trajectory transfer. Distributed storage optimization is not
    its focus, so this baseline still reads data from TQ's
    sample-level storage. Since the original code is not open-sourced,
    our implementation is based on the paper's description.
\end{itemize}

\begin{figure}[t]
  \centering
    \includegraphics[width=0.88\linewidth]{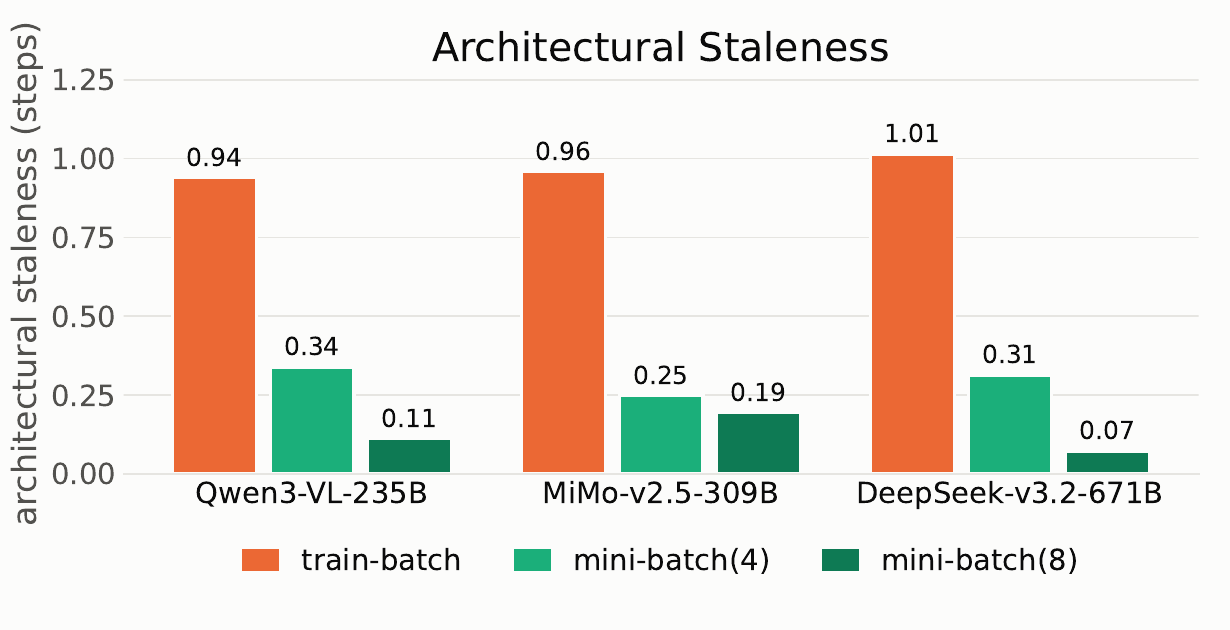}
  \par\vspace{-0.2in}
  \caption{\label{fig:arch-staleness} Architectural staleness of the
    disaggregated system on the CodeAgent task.}
  \vspace{-0.1in}
\end{figure}

\parabf{Testbed and Other Settings.}
All experiments run on a GPU cluster of 32 nodes, each equipped with 8 NVIDIA Hopper-
architecture GPUs and interconnected via high-bandwidth
NVLink and RDMA networks.  
We use GRPO~\cite{shao2024deepseekmath} with $n=16$
rollouts per prompt as the RL algorithm, together with R3~\cite{r3}, which replays the
routing decisions of the rollout engine during training to reduce the
training-inference discrepancy. We filter out samples whose staleness
exceeds an upper bound to limit the parameter-version gap between
generation and training. We also enforce a partial-rollout completion
quota---a minimum completion ratio for every rollout step---to
mitigate long-tail starvation. For
rollout-training placement, every module in the colocated setting occupies all the
GPUs. In the disaggregated setting, we allocate separate GPU
pools to balance their durations as closely as possible, using a
3:1 rollout-to-training GPU split in the CodeAgent experiments.

\begin{figure*}[t]
  \centering
    \includegraphics[width=0.72\textwidth]{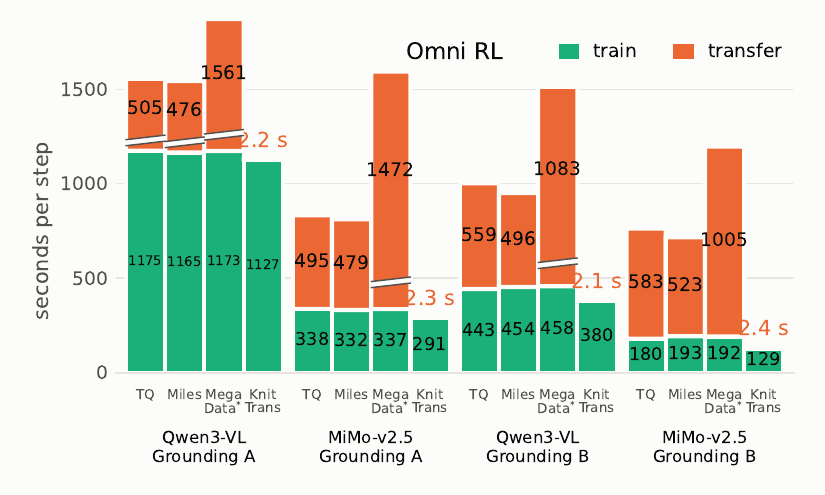}
  \par\vspace{-0.2in}
  \caption{\label{fig:transfer-latency} Exposed transfer and training
    computation latency per step on Omni RL.}
  \vspace{-0.15in}
\end{figure*}

\subsection{End-to-End Performance}
\label{sec:eval-e2e}

\parabf{Architectural Staleness.} We examine whether
mini-batch admission narrows the architectural staleness gap between
the two architectures. We use the CodeAgent task, which exhibits 
pronounced long-tailed rollout durations, with the staleness bound set 
to 4 in all runs. For each
model, we run the disaggregated system with train-batch admission,
i.e., batch-level streaming, and with
\sysname's mini-batch admission at 4 and 8 mini-batches per train
batch. Architectural staleness is measured as the average difference
between the per-step staleness of each run and that of the colocated
run under the same configuration, taken over the steps after the
staleness has stabilized (step $>$ staleness bound). 

Figure~\ref{fig:arch-staleness} shows that mini-batch admission substantially
reduces architectural staleness on all three models. 
On Qwen3-VL-235B, architectural
staleness falls from 0.94 steps under train-batch admission to 0.34 at
4 mini-batches and 0.11 at 8; on MiMo-V2.5-309B, it falls from 0.96 to
0.25 and 0.19; and on DeepSeek-v3.2-671B, from 1.01 to 0.31 and 0.07,
respectively. 
Deviations from the idealized results (0.25 at 4 mini-batches 
and 0.125 at 8 mini-batches) are consistent with per-step variation 
in rollout and training durations, which prevents perfectly balanced overlap.
These results demonstrate the effectiveness of finer-grained admission. 
Four mini-batches per train batch already capture most of the benefit; 
increasing the number to eight yields
diminishing returns without eliminating architectural staleness.

\parabf{Transfer Latency.} We then measure how much \sysname
shortens the rollout-to-training path. To compare this path across
both architectures despite their different rollout-training overlap,
we measure transfer latency exposed on the training critical path
and training computation time per step, comparing \sysname with
the three baselines introduced in \S\ref{sec:eval-setup}.
Figure~\ref{fig:transfer-latency} reports all four configurations on
both Omni RL models and datasets, averaged over the steps after warmup.

Across all four model--dataset pairs, \sysname reduces exposed
transfer latency by $217.6$--$264.3\times$, a reduction of
two orders of magnitude, relative to TQ, leaving training
computation as the dominant cost. The redistribution of multimodal
inputs through collective communication takes about 380--520\,s per
step, accounting for roughly 76--90\% of TQ's exposed transfer latency
across these workloads. The transfer-plus-training
speedup is $1.5$--$5.8\times$. Miles remains
close to TQ in exposed transfer latency, mainly because of
the large number of small tensors in trajectory data: transferring
fragments individually incurs repeated memory-registration and
Store-operation overheads~\cite{mooncake2026miles}.
Meanwhile, changing the transport
backend alone does not remove the aforementioned main overhead. 
The adapted MegaData$^{*}$ baseline suffers from
extensive redundant reads caused by sample-level storage, leading to
severe congestion in distributed storage and stragglers among
training workers that independently fetch trajectory data.
This implies that the storage design---specifically, atomic-unit
storage---remains essential even with topology-aware preprocessing.

\begin{figure}[t]
  \centering
    \includegraphics[width=0.88\linewidth]{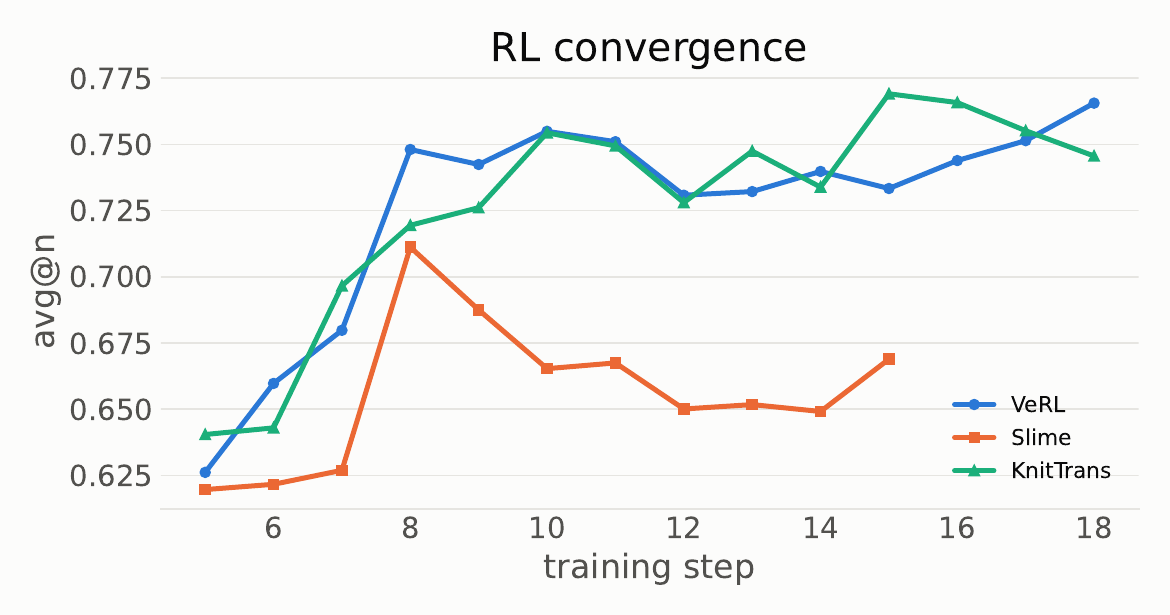}
  \par\vspace{-0.2in}
  \caption{\label{fig:rl-convergence} RL convergence: avg@$n$ per
    training step on the CodeAgent task with MiMo-V2.5.}
  \vspace{-0.2in}
\end{figure}

\begin{figure*}[t]
  \centering
    \begin{minipage}[c]{0.68\linewidth}
      \centering
      \includegraphics[width=\linewidth,trim=0 4bp 0 8bp,clip]{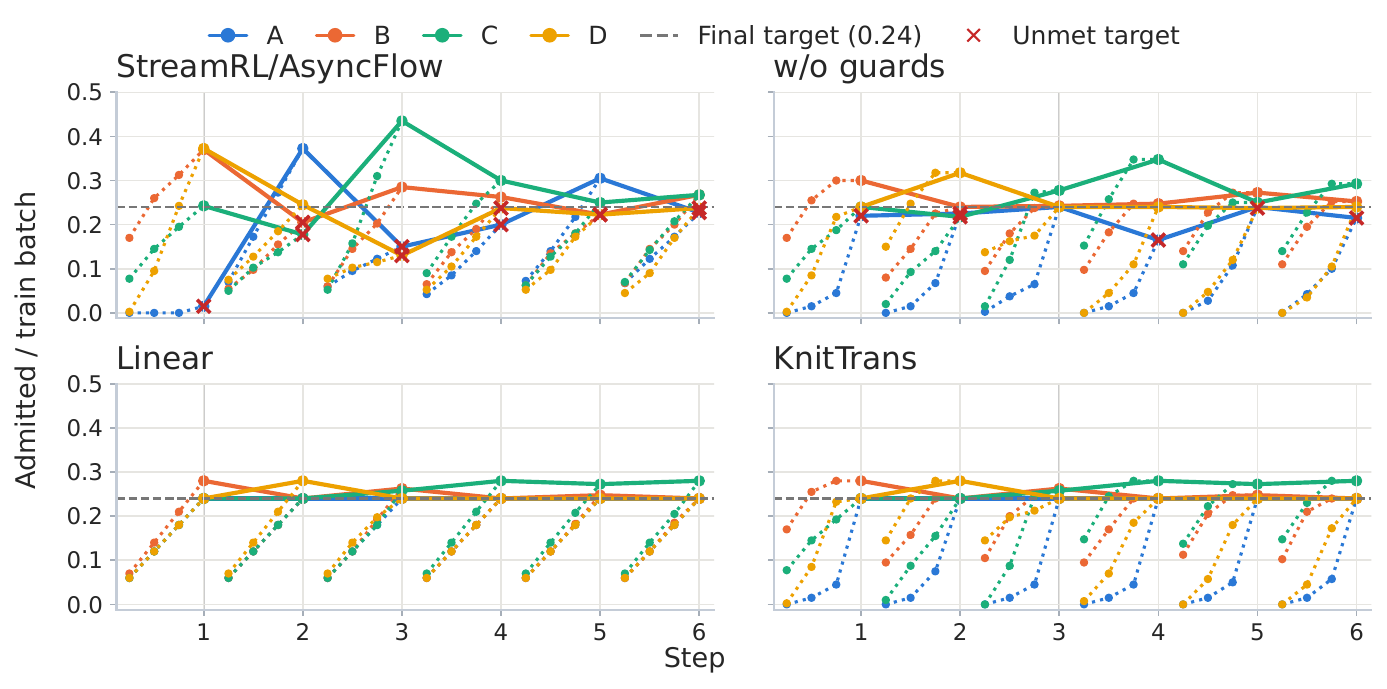}
    \end{minipage}%
    \hspace{0.02\linewidth}%
    \begin{minipage}[c]{0.30\linewidth}
      \centering
      \includegraphics[width=\linewidth]{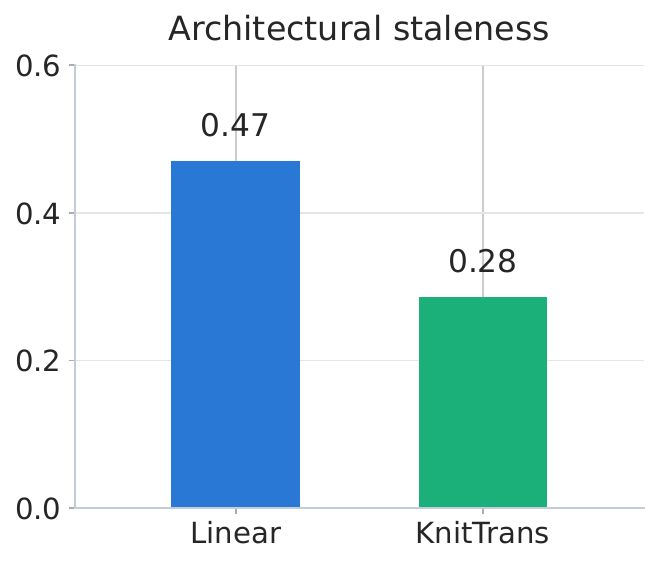}
    \end{minipage}\\[0pt]
    \begin{minipage}[t]{0.68\linewidth}
      \centering\small (a) Per-source batch composition.
    \end{minipage}%
    \hspace{0.02\linewidth}%
    \begin{minipage}[t]{0.30\linewidth}
      \centering\small (b) Architectural staleness.
    \end{minipage}
  \par\vspace{-0.12in}
  \caption{\label{fig:rl-quality} Constraint preservation and architectural
    staleness under different admission policies.}
  \vspace{-0.1in}
\end{figure*}

These results highlight the importance of matching storage units to the
training topology, rather than optimizing transport or preprocessing
in isolation. The evaluation results also show that \sysname
reduces training time through Multimodal Encoding Deduplication.
The combined benefit is larger on Grounding B,
where transfer and encoding
account for more of the rollout-to-training path. Conversely, the
gain is more limited on Qwen3-VL Grounding A, where backbone computation
already dominates due to its full-attention architecture.
The ablations in \S\ref{sec:eval-ablation}
separate the contributions of these mechanisms.


\subsection{RL Quality}
\label{sec:eval-quality}
We evaluate \sysname's impact on RL quality from two perspectives:
\emph{RL Convergence}, which assesses how reduced architectural
staleness affects learning, and \emph{Constraint Preservation}, which
tests whether mini-batch admission preserves algorithm-defined
train-batch constraints while retaining the benefits of streaming.

\parabf{RL Convergence.} \S\ref{sec:eval-e2e} showed that
mini-batch admission reduces architectural staleness; we
further examine its effect on RL convergence.
We conduct experiments on MiMo-V2.5 using three configurations: the
colocated system (VeRL), the disaggregated system with train-batch
admission (slime), and \sysname with 4 mini-batches per train
batch. All other settings remain unchanged. We report avg@$n$
once staleness has stabilized.
Figure~\ref{fig:rl-convergence} reports the curves. \sysname
closely follows the colocated run, with a mean gap of at most 0.007. 
Train-batch admission instead trails the colocated arm by
0.082 avg@$n$ on average. Reducing architectural staleness from 0.96
to 0.25 steps thus narrows the learning-quality gap between the
disaggregated and colocated systems.

\parabf{Constraint Preservation.}
We test the trade-off in \S\ref{sec:design-admission} using a
trace-driven admission replay. All policies use the same source-local
rollout durations and acceptance decisions. We use the four data
sources A--D in Figure~\ref{fig:admission-skewness}(a), each with a target
ratio of 24\%.
\textbf{StreamRL/AsyncFlow} admits accepted samples on completion
without enforcing source quotas.
\textbf{Linear} uses linear interpolation to set progressive
per-source targets.
\textbf{w/o guards} instead enforces the current
cumulative per-source targets, which are initially relaxed to allow
earlier admission and tighten across mini-batches.
\textbf{\sysname} further adds feasibility guards to these targets.

Figure~\ref{fig:rl-quality}(a) shows batch composition during
admission. Without considering batch-wide constraints,
\textbf{StreamRL/AsyncFlow} violates a final quota in every step.
\textbf{w/o guards} also incurs violations: admitting mini-batches
once the current targets are met allows faster sources to
occupy too many slots early.
Linear and \sysname satisfy all final quotas,
but differ in streaming benefits.
During trace replay, we measure architectural staleness
under these two mini-batch admission policies, as shown in
Figure~\ref{fig:rl-quality}(b).
\sysname lowers the average architectural staleness
from 0.470 to 0.286, a 39.2\% reduction relative to Linear.
Relaxed targets with feasibility guards preserve final quotas while
enabling earlier admission than linear targets.

\begin{figure*}[t]
  \centering
    \begin{minipage}[c]{0.33\linewidth}
      \centering
      \includegraphics[width=\linewidth,trim=0 12bp 0 18bp,clip]{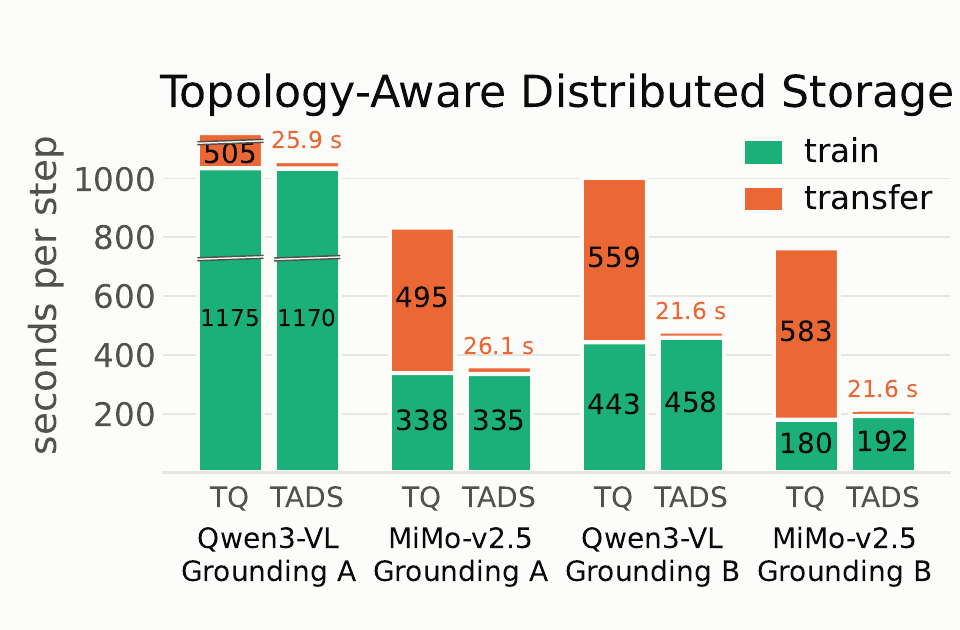}
    \end{minipage}%
    \begin{minipage}[c]{0.33\linewidth}
      \centering
      \includegraphics[width=\linewidth,trim=0 12bp 0 18bp,clip]{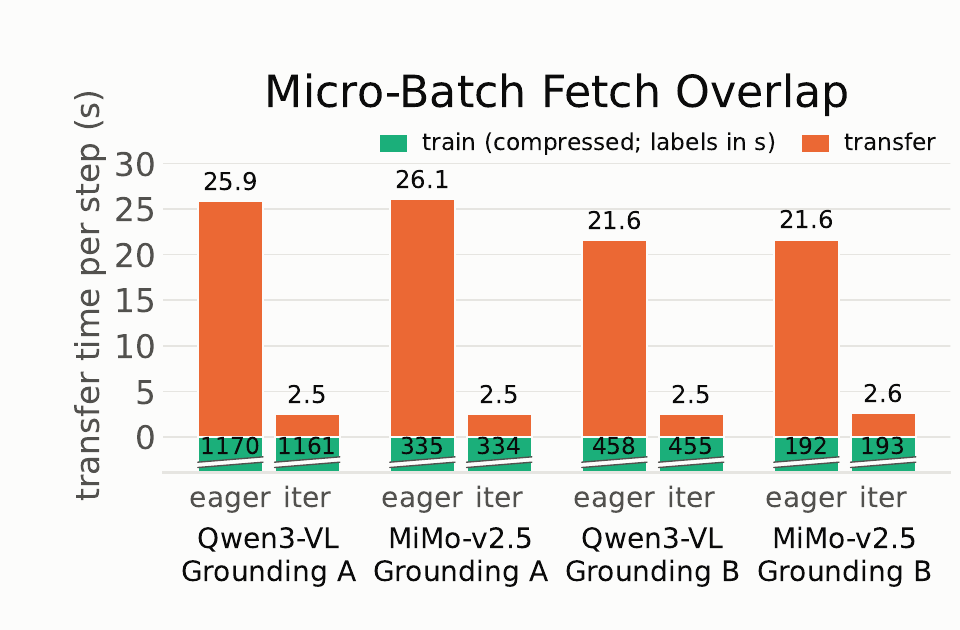}
    \end{minipage}%
    \begin{minipage}[c]{0.33\linewidth}
      \centering
      \includegraphics[width=\linewidth,trim=0 12bp 0 18bp,clip]{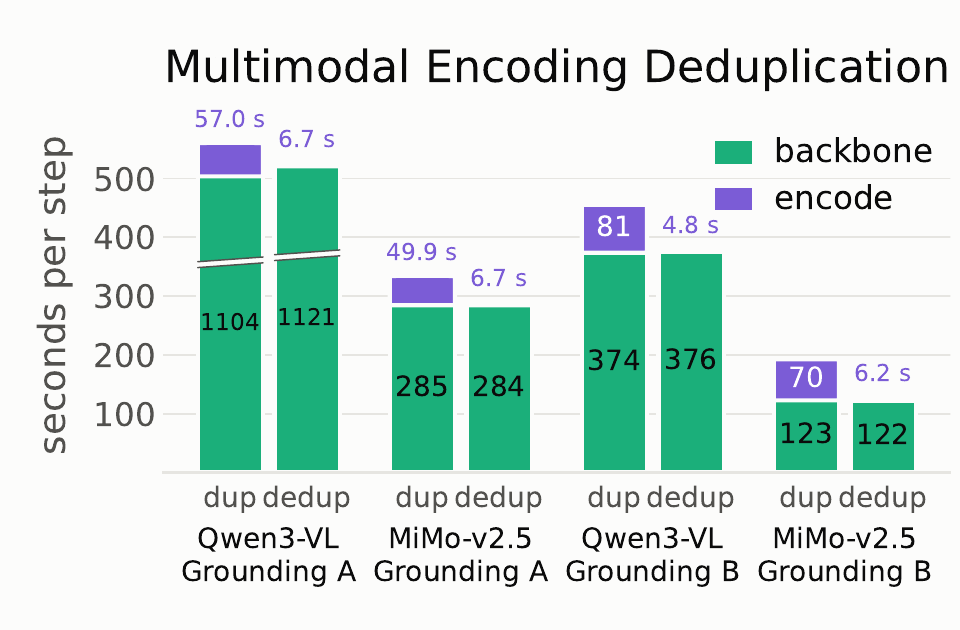}
    \end{minipage}\\[1pt]
    \makebox[0.33\linewidth]{\small (a)}%
    \makebox[0.33\linewidth]{\small (b)}%
    \makebox[0.33\linewidth]{\small (c)}
  \par\vspace{-0.12in}
  \caption{\label{fig:ablation} Layer-wise latency ablations:
    (a)~sample-level \texttt{TQ} versus topology-aware distributed storage (\texttt{TADS});
    (b)~\texttt{eager} versus \texttt{iter} fetching on TADS;
    (c)~duplicated (\texttt{dup}) versus deduplicated (\texttt{dedup}) multimodal encoding
    with iterator fetching. The \texttt{eager} configuration is \texttt{TADS} from (a),
    and \texttt{dup} is \texttt{iter} from (b).}
  \vspace{-0.1in}
\end{figure*}

\subsection{Ablation Study}
\label{sec:eval-ablation}

We examine topology-aware storage and the two training-side
optimizations on the Omni RL task in the colocated setting, using
both multimodal models on Grounding A and Grounding B. Each
pair replays the same traced training batches. We report the mean
over the steps after the warmup.

\parabf{Topology-Aware Distributed Storage.}
Figure~\ref{fig:ablation}(a) compares \texttt{TQ} with
\texttt{TADS}, which organizes trajectory data, especially multimodal
inputs, into independent atomic units and fetches them according to
the training topology, so each rank reads exactly the multimodal
units it consumes.
Eager fetching and duplicated encoding remain unchanged.
Across the four model--dataset pairs, exposed transfer falls from
495--583\,s to 21.6--26.1\,s per step, reducing total
transfer-plus-training time by 28.8--72.0\%.
These gains come from reducing redundant fetches and redistributions of
multimodal inputs, without relying on micro-batch fetch overlap.

\parabf{Micro-Batch Fetch Overlap.}
Figure~\ref{fig:ablation}(b) holds topology-aware storage and the
encoding policy fixed, comparing \texttt{eager} materialization of
the mini-batch with \texttt{iter} fetching and background prefetch.
Exposed transfer falls from 21.6--26.1\,s to 2.5--2.6\,s per step
across the four configurations, an additional
$8.4$--$10.4\times$ reduction over TADS alone.
The remaining exposed latency includes metadata-based scheduling,
cluster-wide dispatch of the resulting plans, and reorganization of
the first micro-batch.
Fetching upcoming micro-batches during computation removes most of the
remaining data wait from the critical path.

\parabf{Multimodal Encoding Deduplication.}
Figure~\ref{fig:ablation}(c) compares \texttt{dup} and \texttt{dedup}
with iterator fetching enabled in both. The latter encodes each unique
multimodal input once per mini-batch and reuses its encoding across
trajectories of the same sample.
With other configurations remaining the same, encoder
forward-plus-backward time falls from 49.9--81.5\,s to 4.8--6.7\,s
per step, corresponding to a 4.3--33.2\% reduction in total
training time. Gains are larger on Grounding B and on MiMo-V2.5, where the
encoder accounts for a greater share of training time.

\section{Related Work}
\label{sec:related}

\parabf{RL Streaming and Training Admission.}
Prior systems refine streaming granularity to overlap rollout
and training in on-policy
settings~\cite{jiang2026flexmarl,chen2026rolloutpipe}.
StreamRL~\cite{zhong2025streamrl} and AsyncFlow~\cite{han2025asyncflow}
support fine-grained streaming in asynchronous rollout--training
pipelines. Batch-wide requirements introduce an additional
restriction on admission: mixed RL enforces target data-source
ratios through per-domain quotas~\cite{datamix,glm2026glm5},
and waiting for all quotas to be met delays otherwise ready
trajectories. \sysname addresses how to admit mini-batches before all
train-batch conditions hold without making the final quotas
unattainable.

\parabf{Trajectory Storage and Data Access.}
Earlier RL data systems support trajectory exchange through
distributed experience storage~\cite{cassirer2021reverb,wang2023gear}.
For LLM RL, TransferQueue~\cite{han2025asyncflow,transferqueue2026}
and Miles~\cite{mooncake2026miles} provide sample-level storage
for rollout-to-training transfer. Beyond RL,
MegaScale-Data~\cite{zhao2026megascale} coordinates distributed
preprocessing and topology-aware data placement, while
BatchWeave~\cite{sun2026batchweave} addresses consistent batch
publication and recovery over object storage.
\sysname extends sample-level exchange with independently
retrievable units matched to consumer-specific assignments.
Unlike pre-training loaders operating on existing datasets,
it coordinates preparation with online trajectory admission
and subsequent micro-batch consumption.

\parabf{Multimodal Training Optimization.}
Prior multimodal training systems address heterogeneous encoder
and backbone workloads through data balancing
~\cite{zheng2025orchmllm}, execution scheduling
~\cite{feng2024optimus}, and disaggregated
training~\cite{zhang2024disttrain}.
Repeated RL trajectories introduce another source of redundant
work: the same multimodal input is consumed across multiple
backbone forward and backward passes.
\sysname addresses how to share its encoder computation 
while preserving every consuming trajectory's gradient
contribution.

\section{Conclusion}
\label{sec:conclusion}

We presented \sysname, a coordinated trajectory-transfer system
for LLM RL. Its constraint-preserving mini-batch admission enables
early training without waiting for full train-batch readiness,
reducing architectural staleness in off-policy
disaggregated execution. On-demand trajectory data access matches
storage units and incremental delivery to each consumer's assignments
and micro-batch schedule. Multimodal encoding deduplication supports
mini-batch-wide reuse while preserving gradient contributions across
backbone backward passes. Experiments 
demonstrate preserved source quotas, a narrower learning-quality
gap with colocated training, and reduced data-transfer and encoding
costs. These results show that effective trajectory transfer requires
coordinating admission and consumption, rather than merely moving
completed trajectories faster.

\clearpage
\bibliography{references}
\bibliographystyle{ACM-Reference-Format}

\end{document}